\documentclass[a4paper, 10 pt, conference]{ieeeconf}  % Comment this line out if you need a4paper

\usepackage{cite}

\usepackage{graphics} % for pdf, bitmapped graphics files
\usepackage{epsfig} % for postscript graphics files
\usepackage{mathptmx} % assumes new font selection scheme installed
\usepackage{times} % assumes new font selection scheme installed
\usepackage{amssymb}  % assumes amsmath package installed

\usepackage{amsmath}

\IEEEoverridecommandlockouts                              
	\title{\Huge Computationally Efficient Laplacian C{\Large{L}}-colME}

\author{Nikola Stankovic, \it{Student Member, IEEE}
\thanks{Podgorica, Montenegro}
}
\usepackage{eso-pic}
\newcommand\AtPageUpperMyright[1]{\AtPageUpperLeft{%
		\put(\LenToUnit{0.5\paperwidth},\LenToUnit{-1cm}){%
			\parbox{0.5\textwidth}{\raggedleft\fontsize{9}{11}\selectfont #1}}%
}}%
\newcommand{\conf}[1]{%
	\AddToShipoutPictureBG*{%
		\AtPageUpperMyright{#1}
	}
}
\begin{document}
\pubid{\makebox[\columnwidth]{979-8-3315-9817-4/26/\$31.00 \copyright 2026 IEEE \hfill} \hspace{\columnsep}\makebox[\columnwidth]{ }}

\maketitle
\conf{30th International Conference on Information Technology (IT) \v{Z}abljak, 24 -- 28 February, 2026}
\pagestyle{empty}

%%%%%%%%%%%%%%%%%%%%%%%%%%%%%%%%%%%%%%%%%%%%%%%%%%%%%%%%%%%%%%%%%%%%%%%%%%%%%%%%
\begin{abstract}

Decentralized collaborative mean estimation (colME) is a fundamental task in heterogeneous networks. Its graph-based variants B-colME and C-colME achieve high scalability of the problem. This paper evaluates the consensus-based C-colME framework, which relies on doubly stochastic averaging matrices to ensure convergence to the oracle solution. We propose CL-colME, a novel variant utilizing Laplacian-based consensus to avoid the computationally expensive normalization processes. Simulation results show that the proposed CL-colME maintains the convergence behavior and accuracy of C-colME while improving computational efficiency.

\end{abstract}

%%%%%%%%%%%%%%%%%%%%%%%%%%%%%%%%%%%%%%%%%%%%%%%%%%%%%%%%%%%%%%%%%%%%%%%%%%%%%%%%

	\section{Introduction}

The widespread use of personal digital devices, together with advances in sensing, computation, and communication technologies, has driven the growth of decentralized, data-driven systems. Agents such as mobile devices, sensors, and IoT platforms continuously generate large amounts of data that are often sensitive and heterogeneous. In these settings, centralized data collection is frequently impractical due to privacy, communication, and scalability constraints. While local processing is possible, learning in isolation usually leads to slow convergence, especially in online scenarios where data arrive sequentially.

These challenges have motivated the development of decentralized and collaborative learning methods. Federated Learning (FL) is a prominent framework in which agents collaboratively learn from local data without sharing raw information \cite{kairouz2021advances,tan2022towards}. Classical FL approaches focus on learning a single global model. However, in many applications, data distributions differ significantly across agents, making a single global solution suboptimal. This has led to increased interest in personalized and clustered FL methods, where agents with similar data distributions collaborate more closely to improve accuracy and convergence speed \cite{ghosh2020efficient,fallah2020personalized,sattler2021clustered,li2020federated,li2021ditto,marfoq2021federated,ding2022collaborative}.

A common approach in personalized FL is to group agents into clusters and train separate models for each group \cite{ghosh2020efficient,sattler2021clustered,ding2022collaborative}. Since agents’ optimal models are unknown in advance, clustering and learning must be performed at the same time. Existing methods typically rely on empirical similarity measures between local model updates \cite{ghosh2020efficient,sattler2021clustered} or assume partial prior knowledge of agent relationships \cite{ding2022collaborative,even2022sample}. In online settings, limited early data can lead to inaccurate similarity estimates and slower convergence.

Mean estimation in decentralized environments is a building block for many collaborative learning problems \cite{dorner2024incentivizing,tsoy2024provable,grimberg2021optimal}. In this context, Collaborative Mean Estimation (colME) was introduced in \cite{asadi2022collaborative} as an online, decentralized framework in which agents exchange information with neighbors to improve their local mean estimates. Subsequent work has improved scalability by using graph-based interaction and consensus mechanism \cite{galante2025scalable}.

A compact-message variant, called B-colME, has been proposed, in which agents exchange aggregated statistics over limited neighborhoods defined by random graphs. While B-colME reduces communication costs, it introduces additional design parameters and may converge more slowly when neighborhood sizes are constrained. Another graph-based approach is the consensus-based colME (C-colME), where agents exchange their current estimates with neighbors and apply a normalized consensus update that ensures equal weighting of information within each connected component. While this approach gives favorable theoretical properties, the construction and normalization of doubly stochastic matrices require division operations at every iteration and for every agent, which can be a computationally demanding  in large-scale or dense networks.

This paper addresses the computational efficiency of C-colME by proposing a Consensus Laplacian-Based variant, termed C{\scriptsize{L}}-colME.\cite{masterwithcode} Rather than explicitly constructing a doubly stochastic averaging matrix, C{\scriptsize{L}}-colME replaces the standard consensus step with a Laplacian-based smoothing update derived directly from the interaction graph among agents. This modification preserves the fundamental structure and objectives of C-colME, namely unbiased collaborative estimation and convergence to class-oracle means within each connected component with minimum variance of error in Gaussian environments. At the same time, it avoids explicit normalization and significantly reduces the per-iteration computational cost. The resulting update admits a simple interpretation as a gradient-based consensus step, which progressively enforces agreement among agents belonging to the same similarity class.

The remainder of the paper is organized as follows. Section II introduces the problem formulation and the collaborative mean estimation framework. Section III presents the proposed Laplacian-based consensus variant C{\scriptsize{L}}-colME and analyzes its convergence properties. Section IV reports numerical results illustrating the efficiency and scalability of the proposed approach. Section V concludes the paper.

\section{Problem Formulation}

We consider a decentralized system consisting of $A=N$ agents, indexed by the set
\[
\mathcal{A}=\{1,2,\dots,A\},
\]
that operate over discrete time $t=1,2,\dots$. Each agent receives streaming data, performs local processing, and can exchange information with a subset of other agents. The key challenge here is to enable agents to identify online which other agents observe statistically similar data, and to exploit this structure for faster and more accurate estimation.

Each agent $a\in\mathcal{A}$ observes at time $t$ a scalar random sample
\[
x_a(t)\in\mathbb{R},
\]
drawn independently from their own distribution $D_a$ with expected value
\[
\mu_a = \mathbb{E}\{x_a(t)\}.
\]
We assume that samples are independent across time and agents, but that the distributions $D_a$ may differ across agents. The vector of observations at time $t$ is denoted by
\[
\mathbf{x}(t)=
\begin{bmatrix}
x_1(t) & x_2(t) & \cdots & x_A(t)
\end{bmatrix}^T.
\]

\smallskip
\noindent\textbf{Similarity classes:}
Agents are said to belong to the same \emph{similarity class} if their data distributions share the same mean. Formally, two agents $a$ and $a'$ belong to the same class if
\[
\Delta_{aa'} \triangleq |\mu_a-\mu_{a'}|=0.
\]
The similarity class of agent $a$ is denoted by $\mathcal{C}_a$. We assume that the set of agents is partitioned into $C$ separate similarity classes, but neither the number of classes which agent belongs to which class are known to the agents.

\smallskip
\noindent\textbf{Objective:}
The goal of each agent $a$ is to estimate its own mean $\mu_a$ as accurately and as quickly as possible. If agent $a$ knew its similarity class $\mathcal{C}_a$, it could collaborate with all agents in that class and significantly reduce estimation error by pooling information. However, since similarity classes are unknown, agents must learn collaboration structure online, which is based on noisy observations. This problem is referred to as \emph{Collaborative Mean Estimation (colME)}.

\subsection{Local Means}

A natural baseline estimator for agent $a$ is its empirical (local) mean, defined after $t$ samples as
\begin{equation}
\bar{x}_{a,a}(t)=\frac{1}{t}\sum_{\tau=1}^{t}x_a(\tau)
=\frac{1}{t}x_a(t)+\frac{t-1}{t}\bar{x}_{a,a}(t-1).
\end{equation}
This estimator converges almost surely to $\mu_a$ as $t\to\infty$, but convergence can be slow, especially in online settings.

Collecting all local means yields the vector
\begin{equation}
\mathbf{X}(t)=
\begin{bmatrix}
\bar{x}_{1,1}(t) &  \cdots & \bar{x}_{A,A}(t)
\end{bmatrix}^T=\frac{1}{t}\mathbf{x}(t)+\frac{t-1}{t}\mathbf{X}(t-1).
\end{equation}

An equivalent formulation, often used in distributed implementations, is based on local sums
\begin{equation}
m_{a,a}(t)=\sum_{\tau=1}^{t}x_a(\tau)=t\,\bar{x}_{a,a}(t).
\end{equation}

\smallskip
\noindent\textbf{Oracle benchmark:}
If similarity classes were known in advance, the optimal (oracle) estimate for agent $a$ at time $t$ would be obtained by averaging local means over all agents in $\mathcal{C}_a$. This oracle solution represents the best achievable performance and will be used as a reference in the results.

\subsection{Confidence Intervals and Similarity Detection}

To learn similarity classes online, agents rely on confidence intervals constructed around their local means. For each agent $a$, define the confidence interval
\begin{equation}
\mathbb{I}_a(t)=
\big[\bar{x}_{a,a}(t)-\beta_\delta(t),\;
\bar{x}_{a,a}(t)+\beta_\delta(t)\big],
\end{equation}
where $\beta_\delta(t)$ is chosen such that
\begin{equation}
\mathbb{P}\Big(|\bar{x}_{a,a}(t)-\mu_a|>\beta_\delta(t)\Big)\le \delta,
\qquad
\lim_{t\to\infty}\beta_\delta(t)=0.
\end{equation}
This guarantees that the true mean $\mu_a$ lies within $\mathbb{I}_a(t)$ with probability at least $1-\delta$, and that confidence intervals shrink over time.
For instance, if observations are $\sigma$-sub-Gaussian, a Laplace-type bound yields \cite{asadi2022collaborative,galante2025scalable}
\begin{equation}
\beta_{\delta}(t)
=\sigma \sqrt{\frac{2}{t}\Big(1+\frac{1}{t}\Big)
\ln\Big(\frac{\sqrt{t+1}}{\delta/2}\Big)}.
\end{equation}

\subsection{Graph-Based Collaboration and Pruning}

In  the initial colME algorithm all agents were communicating with each other, which resulted in scalability problems for large graphs. In graph-based B-colME and C-colME the agents communicate only with neighbors defined by the initial random graph, with maximum degree $r$.  Initially, agents may collaborate with neighbors belonging to different similarity classes. At each time instant $t$, agent $a$ compares its confidence interval with those of its neighbors. An edge between agents $a$ and $a'$ is retained if and only if
\begin{equation}
\mathbb{I}_a(t)\cap\mathbb{I}_{a'}(t)\neq \emptyset.
\end{equation}
If the intervals do not intersect, the edge is removed. This mechanism exploits two key properties: confidence intervals of agents in the same similarity class continue to intersect with high probability, while intervals of agents from different classes eventually separate as time $t$ increases.
 As a result, $\mathbf{A}(t)$ evolves over time by pruning unlikely collaborations, meaning that agents interact over a time-varying graph described by a time-dependent adjacency matrix $\mathbf{A}(t)$.

\subsection{Expected Separation Time}

Consider two agents $a$ and $b$ from different similarity classes, with mean gap $\Delta=|\mu_a-\mu_b|>0$. A rough estimate of the time required for their confidence intervals to separate is obtained by solving
$
2\,\beta_\delta(t)\approx \Delta.$ 

In systems with multiple classes, the overall separation time is governed by the smallest inter-class gap
$\min_{a\neq b} \Delta_{ab}$.

After this separation occurs, the collaboration graph stabilizes and meaningful collaborative estimation can proceed.

\subsection{Consensus-based Approach C-colME}

The consensus-based approach (C-colME) uses both the local means $\bar{x}_{a,a}(t)= \frac{1}{t}\sum_{\tau=1}^{t}x_a(\tau)$, or in vector form, for all agents $\mathbf{X}(t)$, and collaborative means calculated using the estimated similarity class, within the random neighborhood $\boldsymbol{\mu}(t)$.

We start with $\boldsymbol{\mu}(0)= \mathbf{X}(0)$ and calculate all next consensus values combining local means $\mathbf{X}(t)$, with weight $(1-\alpha(t))$ and previous consensus mean $\boldsymbol{\mu}(t-1)$ with weight $\alpha(t)$, that is
\begin{equation}\boldsymbol{\mu}(t+1)=(1-\alpha(t))\mathbf{X}(t)+\alpha(t)\mathbf{W}(t)\boldsymbol{\mu}(t)\end{equation}
where $\mathbf{W}(t)$ is the weight matrix that will be discussed next.

\begin{itemize}
	\item
	In order to obtain an unbiased result, we must obviously have the sum of the rows of the matrix $\mathbf{W}_t$ to be one. If we assume that all means are equal to 1, then $\boldsymbol{1}=(1-\alpha(t))\boldsymbol{1}+\alpha(t)\mathbf{W}(t)\boldsymbol{1}$, should hold, which means that all rows of $\mathbf{W}$ should sum to 1. Since this matrix is symmetric, this holds for columns as well, and such a matrix is called a double stochastic matrix. 
	
	\item At the same time, the matrix $\mathbf{W}(t)$ should play the role of the adjacency matrix $\mathbf{A}(t)$ (weighted adjacency matrix) since it should combine the values of $\boldsymbol{\mu}(t-1)$ within the neighborhood of a given node only, hoping that this neighborhood would be a part of the similarity class (after some time steps).
	
	\item We will assume that the values of matrix $\mathbf{W}(t)$ are constant for each node, and take their value as 
	\begin{equation}W_{i,j}(t)=W_{j,i}(t)=\frac{1}{\max \{D_i,D_j\}+1} \text{ for } A_{i,j}(t)=1\end{equation}
	and $W_{i,j}(t)=0$ for $A_{i,j}=0$. The values $D(i)$ are degrees of matrix $\mathbf{A}(t)$, that is $D(i)=\sum_jA_{i,j}(t)$
	
	\item In order to achieve that $\sum_jW_{i,j}(t)=1$ we have 
	\begin{equation}W_{i,i}(t)=1-\sum_jW_{i,j}(t).\end{equation}
	
\end{itemize}

	\section{{Consensus Laplacian-Based Variant} C{\scriptsize{L}}-colME}
	
	Consider the data set $x_a(t)$ in the graph $\mathcal{G}(\mathcal{A})$ with adjacency matrix $\mathbf{A}$. The Laplacian on this graph is defined by $\mathbf{L}=\mathbf{D}-\mathbf{A}$ where $\mathbf{D}$ is the diagonal matrix (called degree matrix) with diagonal elements equal to the number of neighbors of each agent (sum of the row elements in $\mathbf{A}$), that is 
	$D_{aa}=|\mathcal{N}_a|=\sum_jA_{a,j}$. 
	Data $x_a(t)$ smoothness on the graph is defined by
	\begin{equation}\mathcal{L}=\sum_a\sum_j A_{a,j}(x_a(t)-x_j(t))^2=\mathbf{x}^T\mathbf{L}\mathbf{x}\end{equation} 
	The data smoothness is a non-negative function, having minimum $\mathcal{L}=0$ when $x_a(t)=x_j(t)=constant$ for each node $a$ within a connected component of the graph. The minimum corresponds to maximally smooth, that is, constant data.
	
	To achieve a smooth version $\mu_a(t)$ of local means $X_a(t)$, we can use steepest descent approach, moving by small steps in the direction of negative gradient of smoothness,
	$\boldsymbol{\mu}^T\mathbf{L}\boldsymbol{\mu}$,  of $\mu_a(t)$ on graph 
	\begin{equation}\frac{\partial \mathcal{L}}{\partial \boldsymbol{\mu}^T}=2\mathbf{L}\boldsymbol{\mu}(t)) \end{equation}
	or in the standard iterative steepest descent form
	\begin{equation}\boldsymbol{\mu}(t)=\boldsymbol{\mu}(t-1)- \beta \mathbf{L}\boldsymbol{\mu}(t-1)=(\mathbf{I}-\beta\mathbf{L})\boldsymbol{\mu}(t-1)\end{equation}
	with $\boldsymbol{\mu}(0)=\mathbf{X}(0)$ being the initial vector of local means.

	Since we want to include local means for initial period, until the graph is sufficiently separated and wrong connections are excluded, we add local means and use the iterative formula (in consensus form) 
		\begin{equation}\boldsymbol{\mu}(t)=(1-\alpha(t))\mathbf{X}(t)+\alpha(t) (\mathbf{I}-\beta\mathbf{L})\boldsymbol{\mu}(t-1)\end{equation}.
        If the Laplacian is replaced by $\mathbf{L}=\mathbf{D}-\mathbf{A}$  we get
		\begin{equation}\boldsymbol{\mu}(t)=(1-\alpha(t))\mathbf{X}(t)+\alpha(t) (\mathbf{I}-\beta\mathbf{D}+\beta\mathbf{A})\boldsymbol{\mu}(t-1)\end{equation}

    Our analysis will focus on the time when $\alpha(t) \to 1$, that is,  when $\boldsymbol{\mu}(t)=\boldsymbol{\mu}(t-1)- \beta \mathbf{L}\boldsymbol{\mu}(t-1)$. 
Fist we can conclude that if the vector $\boldsymbol{\mu}(t-1)=\boldsymbol{\mu}_a=\mu_a\mathbf{1}$ reached a constant and the same value over all agents then, by the definition of Laplacian, follows $\mathbf{L}\boldsymbol{\mu}(t-1)=\mathbf{L}\boldsymbol{\mu}_a=\mu_a\mathbf{L}\mathbf{1}=0$ and the steady state $\boldsymbol{\mu}(t)=\boldsymbol{\mu}(t-1)=\boldsymbol{\mu}_a$ is reached. 
    
    Assume that at instant $t_0=t-1$ the confidence interval based disconnections are finished and the graph remains constant. Also we assume that by this instant $\alpha(t) \to 1$. 
    Then we can write
    $\boldsymbol{\mu}(t_0+1)= (\mathbf{I}-\beta\mathbf{L})\boldsymbol{\mu}(t_0)$, 
    $\boldsymbol{\mu}(t_0+2)= (\mathbf{I}-\beta\mathbf{L})\boldsymbol{\mu}(t_0+1)= (\mathbf{I}-\beta\mathbf{L})^2\boldsymbol{\mu}(t_0)$, or
    \begin{gather}
    \boldsymbol{\mu}(t_0+n)= (\mathbf{I}-\beta\mathbf{L})^n\boldsymbol{\mu}(t_0)
    \end{gather}
Since the Laplacian is symmetric matrix we can always find its eigenvalue decomposition
\begin{equation}
\mathbf{L}=\mathbf{U}\boldsymbol{\Lambda}\mathbf{U}^T
\end{equation}
The eigenvalue decomposition of matrix $(\mathbf{I}-\beta\mathbf{L})^n$ is
\begin{equation}
(\mathbf{I}-\beta\mathbf{L})^n=\mathbf{U}(\mathbf{I}-\beta\boldsymbol{\Lambda})^n\mathbf{U}^T
\end{equation}
with eigenvalues $1-\beta \lambda_i$, where $\lambda_i$ are eigenvalues of the Laplacian. 	 For the Laplacian holds	$\mathbf{L}\mathbf{1}=\mathbf{0}$. 
	It means than $\lambda_1=0$ is the eigenvalue for a constant eigenvector. All other eigenvalues $\lambda_i$ are greater than 0 since the Laplacian is a real-valued symmetric matrix.  The eigenvector corresponding to the eigenvalue $\lambda_1=0$ is constant with elements $1/N$ where $N$ is the number of nodes in a graph component.

The constant $\beta$ in the gradient descent should be chosen such that   $|\beta \lambda_{max}|<1$, where $\lambda_{max}$ is the maximum eigenvalue of the Laplacian, $\mathbf{L}$.  Note that $\lambda_{min}=0$ by the Laplacian definition, with multiplicity equal to the number of graph components (classes of agents, after wrong links are disconnected, assuming no isolated agents appear). 

With this condition, for notation simplicity assuming only one graph component, we have
\begin{gather}
\lim_{n \to \infty}(\mathbf{I}-\beta\mathbf{L})^n=\lim_{n \to \infty} \mathbf{U}(\mathbf{I}-\beta\boldsymbol{\Lambda})^n\mathbf{U}^T = \nonumber \\
 \lim_{n \to \infty} \begin{bmatrix}\frac{\mathbf{1}}{\sqrt{N}} \! \! & \mathbf{u}_1 \! & \! \dots & \! \mathbf{u}_N  \end{bmatrix} \! \! \begin{bmatrix} 1 \! & 0 & \! \dots \! & 0 \nonumber \\ 
0 \! & 1- \beta \lambda_1 & \! \dots \! & 0 \\
0 \! & 0 & \! \dots \! & 1- \beta \lambda_N\end{bmatrix}^n \! \!
\begin{bmatrix}\frac{\mathbf{1}^T}{\sqrt{N}} \\ \mathbf{u}_1^T \\ \vdots \\ \mathbf{u}^T_N  \end{bmatrix} \nonumber \\
=  \mathbf{U} \begin{bmatrix} 1 & 0 & \dots & 0 \\ 
0 & 0 & \dots & 0 \\
0 & 0 & \dots & 0\end{bmatrix} 
\mathbf{U}^T 
=\begin{bmatrix} \frac{1}{N} & \frac{1}{N} & \dots & \frac{1}{N} \\ 
\frac{1}{N} & \frac{1}{N} & \dots & \frac{1}{N} \\
\frac{1}{N} & \frac{1}{N} & \dots & \frac{1}{N}\end{bmatrix}
\end{gather}
This is exactly the averaging operator for minimum variance estimation of the mean in the case of Gaussian environment. The same holds for each graph component. It means that although the graph is random, within few iterations the solution behaves as all agents were fully connected and the recursion tends toward class oracle solutions. 

\begin{figure}[htbp]
	\centering
	
	\includegraphics[scale=0.7, trim={0 0 5.8cm 0}, clip]{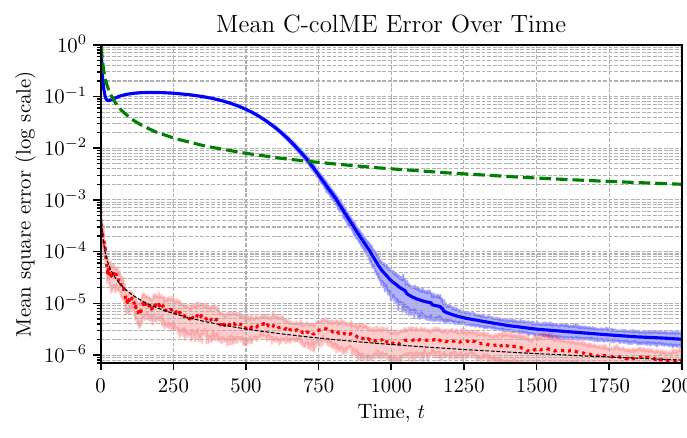}
		
	\includegraphics[scale=0.7, trim={0 0 5.8cm 0}, clip]{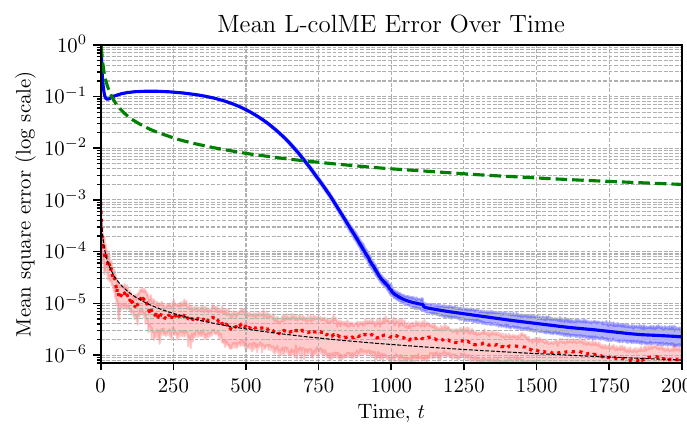}

	\caption{C-colME and C{\scriptsize{L}}-colME ($N=5000$): Total MSE of the estimation averaged over agents at time instants $t$ (blue line). The results are averaged over 10 realizations with $\beta=0.1$.
		\textbf{Calculation time} for C{\scriptsize{L}}-colME is 722s per realization, on average, while for C-colME it was 871s.  Local solution (green dashed line) and oracle solution (red dotted line) are also given.}
	\label{Fig_2L}
\end{figure}

\section{Numerical Example}

We have implemented C-colME and C{\scriptsize{L}}-colME on a system with $N=5000$ agents, with two equal classes whose means are $\mu_1=1.2$ and $\mu_b=2$ and a Gaussian additive noise with standard deviation $\sigma=2$. The simulation is run 10 times and the results for the MSE of the solution are given in Fig. \ref{Fig_2L}, along with the local solution (green dashed line) and oracle solution (red dotted line). 
The convergence rates of C-colME and C{\scriptsize{L}}-colME is almost the same, although the second largest eigenvalues of the matrices $\mathbf{W}_t$ and $\mathbf{I}-\beta\mathbf{L}$ may slightly differ. It has been shown in \cite{galante2025scalable} that the doubly stochastic matrix $\mathbf{W}_t$ yields optimal averaging within each connected component in terms of minimizing estimation error. We have also concluded that C{\scriptsize{L}}-colME produces equivalent of fully connected graph over iterations, resulting in minimum variance of error for Gaussian environments.  Constructing and normalizing $\mathbf{W}_t$ requires division operations at every agent and time instant, which constitutes the most computationally demanding step of C-colME, being avoided in C{\scriptsize{L}}-colME. The step size $\beta$ plays the role of a learning rate in C{\scriptsize{L}}-colME.

\section{Conclusion}

This paper addressed collaborative mean estimation in decentralized and heterogeneous networks, with emphasis on computational efficiency and scalability. We reviewed the consensus-based C-colME framework and its reliance on doubly stochastic averaging matrices to guarantee unbiased estimation and convergence to the oracle solution. A Laplacian-based consensus variant, C{\scriptsize L}-colME, was proposed as a computationally efficient modification that preserves the fundamental structure and objectives of C-colME while avoiding explicit normalization. Numerical results indicate that the proposed approach reduces computation time by approximately 30\% compared to standard C-colME, while exhibiting similar convergence behavior and estimation accuracy. These results suggest that Laplacian-based consensus updates provide an effective alternative for large-scale collaborative mean estimation.

\bibliographystyle{IEEEtran}
\bibliography{reference}

@article{asadi2022collaborative,
  title={Collaborative algorithms for online personalized mean estimation},
  author={Asadi, Mahsa and Bellet, Aur{\'e}lien and Maillard, Odalric-Ambrym and Tommasi, Marc},
  journal={arXiv preprint arXiv:2208.11530},
  year={2022}
}

@inproceedings{galante2025scalable,
  title={Scalable Decentralized Algorithms for Online Personalized Mean Estimation},
  author={Galante, Franco and Neglia, Giovanni and Leonardi, Emilio},
  booktitle={Proceedings of the AAAI Conference on Artificial Intelligence},
  volume={39},
  number={16},
  pages={16699--16707},
  year={2025}
}

@article{kairouz2021advances,
  title={Advances and open problems in federated learning},
  author={Kairouz, Peter and McMahan, H Brendan and others},
  journal={Foundations and Trends{\textregistered} in Machine Learning},
  volume={14},
  number={1--2},
  pages={1--210},
  year={2021}
}

@article{tan2022towards,
  title={Towards personalized federated learning},
  author={Tan, Alysa Ziying and Yu, Han and Cui, Lizhen and Yang, Qiang},
  journal={IEEE Transactions on Neural Networks and Learning Systems},
  volume={34},
  number={12},
  pages={9587--9603},
  year={2022}
}

@inproceedings{ghosh2020efficient,
  author={Ghosh, Avishek and Chung, Jaehoon and Yin, Dong and Ramchandran, Kannan},
  title={An Efficient Framework for Clustered Federated Learning},
  booktitle={Advances in Neural Information Processing Systems},
  year={2020}
}

@inproceedings{fallah2020personalized,
  author={Fallah, Alireza and Mokhtari, Aryan and Ozdaglar, Asuman},
  title={Personalized Federated Learning with Theoretical Guarantees},
  booktitle={Advances in Neural Information Processing Systems},
  year={2020}
}

@article{sattler2021clustered,
  author={Sattler, Felix and M{\"u}ller, Klaus-Robert and Samek, Wojciech},
  title={Clustered Federated Learning},
  journal={IEEE Transactions on Neural Networks and Learning Systems},
  volume={32},
  number={8},
  pages={3710--3722},
  year={2021}
}

@article{li2020federated,
  author={Li, Tian and Sahu, Anit Kumar and Talwalkar, Ameet and Smith, Virginia},
  title={Federated Learning: Challenges, Methods, and Future Directions},
  journal={IEEE Signal Processing Magazine},
  volume={37},
  number={3},
  pages={50--60},
  year={2020}
}

@inproceedings{li2021ditto,
  author={Li, Tian and Hu, Shiqiang and Beirami, Ahmad and Smith, Virginia},
  title={Ditto: Fair and Robust Federated Learning Through Personalization},
  booktitle={International Conference on Machine Learning},
  year={2021}
}

@inproceedings{marfoq2021federated,
  author={Marfoq, Othmane and Neglia, Giovanni and Bellet, Aur{\'e}lien and Kameni, Laetitia and Vidal, Richard},
  title={Federated Multi-Task Learning under a Mixture of Distributions},
  booktitle={Advances in Neural Information Processing Systems},
  year={2021}
}

@inproceedings{ding2022collaborative,
  author={Ding, Shengyu and Wang, Weina},
  title={Collaborative Learning by Detecting Collaboration Partners},
  booktitle={Advances in Neural Information Processing Systems},
  year={2022}
}

@inproceedings{even2022sample,
  author={Even, Maayan and Massouli{\'e}, Laurent and Scaman, Kevin},
  title={On Sample Optimality in Personalized Collaborative and Federated Learning},
  booktitle={Advances in Neural Information Processing Systems},
  year={2022}
}

@inproceedings{dorner2024incentivizing,
  author={Dorner, Florian E. and Konstantinov, Nikolai and others},
  title={Incentivizing honesty among competitors in collaborative learning and optimization},
  booktitle={Advances in Neural Information Processing Systems},
  year={2024}
}

@inproceedings{tsoy2024provable,
  author={Tsoy, Nikolai and Mihalkova, Anna and Todorova, T. N. and Konstantinov, Nikolai},
  title={Provable Mutual Benefits from Federated Learning in Privacy-Sensitive Domains},
  booktitle={International Conference on Artificial Intelligence and Statistics},
  year={2024}
}

@article{grimberg2021optimal,
  title={Optimal model averaging: Towards personalized collaborative learning},
  author={Grimberg, Felix and Hartley, Mary-Anne and Karimireddy, Sai P. and Jaggi, Martin},
  journal={arXiv preprint arXiv:2110.12946},
  year={2021}
}

@book{masterwithcode,
  author    = {N. Stankovic},
  title     = {Variance-Aware ColME: Enhancing Decentralized Online
  Mean Estimation},
  publisher = {Master Thesis, Politecnico di Torino},
  year      = {October, 2025}
}
 
\end{document}